\documentclass[a4paper]{spie}  

\newcommand*\arcsec{\ensuremath{^{\prime\prime}}}
 
\usepackage{amsmath,amsfonts,amssymb}
\usepackage{graphicx}
\usepackage[colorlinks=true, allcolors=blue]{hyperref}
\usepackage{longtable}

\title{The MICADO first-light imager for the ELT\\ First steps of the SCAO system MAIT}

\author[a]{Yann Cl\'enet} 
\author[a]{Eric Gendron} 
\author[a]{Fabrice Vidal}
\author[b]{Mathieu Cohen}
\author[a]{Fr\'ed\'eric Chapron} 
\author[a]{Arnaud Sevin}
\author[a]{Tristan Buey} 
\author[c]{Sylvain Guieu} 
\author[b]{Sylvestre Taburet}
\author[a]{Bruno Borgo}
\author[b]{Jean-Michel Huet}
\author[a]{Olivier Dupuis}
\author[a]{K\'evin Cloiseau}
\author[d]{Alexandre Blin}
\author[b]{Julien Gaudemard}
\author[a]{Claude Collin}
\author[a]{Julian Porras}
\author[a]{Florian Ferreira}
\author[a]{Jordan Raffard}
\author[b]{Fanny Chemla}
\author[a]{Vincent Lapeyr\`ere}
\author[e]{Eric Meyer} 
\author[e]{Nicolas Gautherot} 
\author[e]{Emmanuel Tisserand} 
\author[e]{Herv\'e Locatelli} 
\author[b]{Gilles Fasola}
\author[a]{Lahoucine Ghouchou}
\author[a]{Camille Gennet}
\author[e]{Fran\c{c}ois Meyer} 
\author[a,f]{Amal Zidi} 
\author[f]{Caroline Kulcs\'ar} 
\author[f]{Henri-Fran\c{c}ois Raynaud} 
\author[g]{Benoit Sassolas} 
\author[g]{Laurent Pinard} 
\author[g]{Christophe Michel} 
\author[a]{Damien Gratadour}
\author[a]{Roderick Dembet} 
\author[a]{Pierre Baudoz} 
\author[a]{Elsa Huby} 
\author[h]{Sebastian Rabien}
\author[h]{Eckhard Sturm}
\author[h]{Richard Davies}

\affil[a]{LESIA, Observatoire de Paris, Universit\'e PSL, CNRS, Sorbonne Universit\'e, Universit\'e de Paris Cit\'e, France}
\affil[b]{GEPI, Observatoire de Paris, Universit\'e PSL, CNRS, France}
\affil[c]{EFISOFT, INSU, CNRS, France}
\affil[d]{DT-INSU, CNRS, France}
\affil[e]{OSU THETA, CNRS, Universit\'e Bourgogne Franche Comt\'e, France}
\affil[f]{LCF, IOGS, CNRS, Universit\'e Paris Saclay, France}
\affil[g]{LMA, IP2I Lyon, CNRS, Universit\'e Claude Bernard Lyon 1, France}
\affil[h]{MPE, Germany}

\authorinfo{Further author information: send correspondence to Y. Cl\'enet, e-mail: yann.clenet "at" obspm.fr }

\begin{document} 
\maketitle

\begin{abstract}
MICADO is the ELT first light instrument, an imager working at the diffraction limit of the telescope thanks to two adaptive optics (AO) modes: a single conjugate one (SCAO), available at the instrument first light and developed by the MICADO consortium, and a multi conjugate one (MCAO), developed by the MORFEO consortium. 

Although the project final design review process is about to be completed, the review board and ESO acknowledged that "the review of the final design can be considered complete for the majority of the MICADO sub-systems" and agreed that MICADO can start manufacturing. 

For the MICADO SCAO module, we have started the manufacturing of several parts: the majority of the SCAO optics and of the SCAO mechanics, the real-time computer software and the instrument control software. This manufacturing is ordered in several steps to allow the progressive integration of a first full AO close loop with the final SCAO parts. 

In this contribution, we will focus on the first two steps: on our AO S\'esame bench and the so-called "$\beta$ flat configuration". We will present the  status of this manufacturing and the first results obtained.
\end{abstract}


\section{AN INTRODUCTION TO MICADO}
\label{sec:intro}  

The MICADO imager is the ELT first light instrument \cite{sturm24}. Being available at telescope first light, MICADO will address all key topics of modern astronomy, from Solar system objects to the first galaxies of the Universe and including exoplanets. Working in the near-IR (0.8-2.4 $\mu$m) at the ELT diffraction limit, it will offer four observing modes:

\newpage
\begin{itemize}
\vspace{-3mm}
\setlength{\itemsep}{0pt}
\item Standard imaging: with 1.5 \& 4 mas pixel scales, the corresponding FoV will be 19 \& 51 arcseconds$^2$. More than 30 broad-band \& narrow-band filters will be available.

\item Astrometric imaging: it drives MICADO design, with a gravity invariant implementation, a fixed mirror optical design, state-of-the-art ADC and dedicated
astrometric calibration and data pipeline.

\item High contrast imaging \cite{baudoz24,huby24a,huby24b}: it will use the central detector and will be enabled via a classical configuration of focal plane coronagraphs and Lyot stops, as well as pupil plane vAPP coronagraphs and sparse aperture masking. Pupil tracking will be available for angular differential imaging. 

\item Slit spectroscopy: it will provide coverage of a wide wavelength range simultaneously (J: 1.15-1.35 $\mu$m, HK: 1.52-2.45 $\mu$m or IzJ: 0.85-1.56 $\mu$m) at a resolution of ~20000 on faint compact or unresolved sources. Three slits will be available: 3\arcsec$\times$16 mas (IzJ), 15\arcsec$\times$20 mas (J \& HK), 3\arcsec$\times$48 mas (IzJ \& HK).
\end{itemize}

\begin{figure}[t]
\begin{center}
   \begin{tabular}{c c}
 \includegraphics[height=6cm]{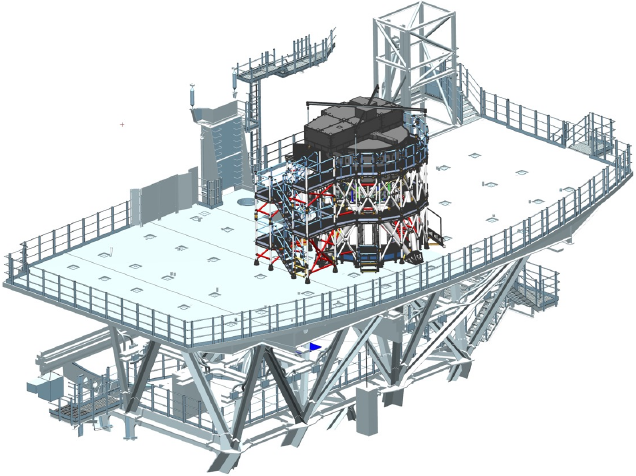} & \includegraphics[height=6cm]{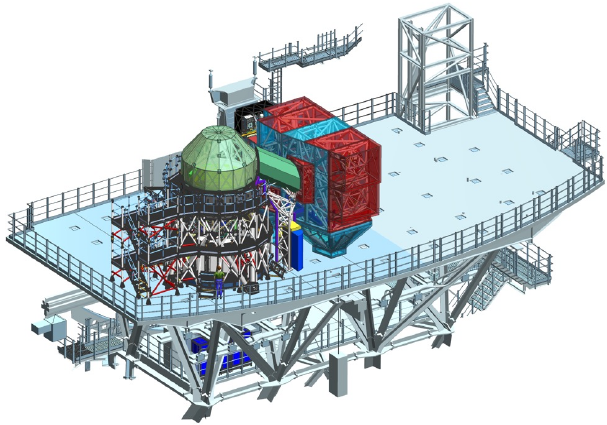}
       \end{tabular}
\end{center}
\caption{\label{fig:micado_standalone} Left: MICADO at the telescope in its standalone configuration, i.e. with the SCAO module alone and a dedicated optical relay. Right: MICADO at the telescope, coupled with MORFEO}
\end{figure}

MICADO will know two successive phases:
\begin{itemize}
\vspace{-3mm}
\setlength{\itemsep}{0pt}
\item the standalone phase (Fig.~\ref{fig:micado_standalone} left), during which it will benefit only from a SCAO correction, developed within the consortium. A passive optical relay is developed by the consortium to feed the instrument. MICADO will be installed on the Nasmyth platform A of the ELT.
\item a so-called "M\&M" phase (Fig.~\ref{fig:micado_standalone} right), after the arrival of MORFEO at the telescope, few years after the ELT and MICADO first light. The instrument will then benefit from both the SCAO and a MCAO correction, the latter being developed by MORFEO together with an active optical relay \cite{ciliegi24}. In this phase, MICADO and MORFEO will be on the ELT Nasmyth platform B, requiring to move MICADO (including SCAO) from platform A to B. The SCAO module is not modified when MORFEO is installed: the space envelop, the interlock system, and the software configuration are  accounting for MORFEO  from the MICADO first light.
\end{itemize}

MICADO current planning is the following:
\begin{itemize}
\vspace{-3mm}
\setlength{\itemsep}{0pt}
\item 11/2018: PDR
\item 04/2021 -  07/2024: FDR sessions
\item 12/2027: PAE
\item end of 2028: commissioning
\item mid 2029: operations
\end{itemize}

\newpage
Early 2023, we have passed  the 4$^{th}$ FDR session. The FDR board report stated after this session that: "the review of the final design can be considered complete for the majority of the MICADO sub-systems, with no showstoppers over the design". The FDR board and ESO agreed then that MICADO  can start procurement and manufacturing. Hence, despite the FDR is not officially passed, MICADO, and in particular the SCAO module, has started the MAIT process. 

\section{THE MICADO SCAO: MAIN SPECIFICATIONS AND FEATURES}
\begin{figure}[t]
\begin{center}
 \includegraphics[height=7.2cm]{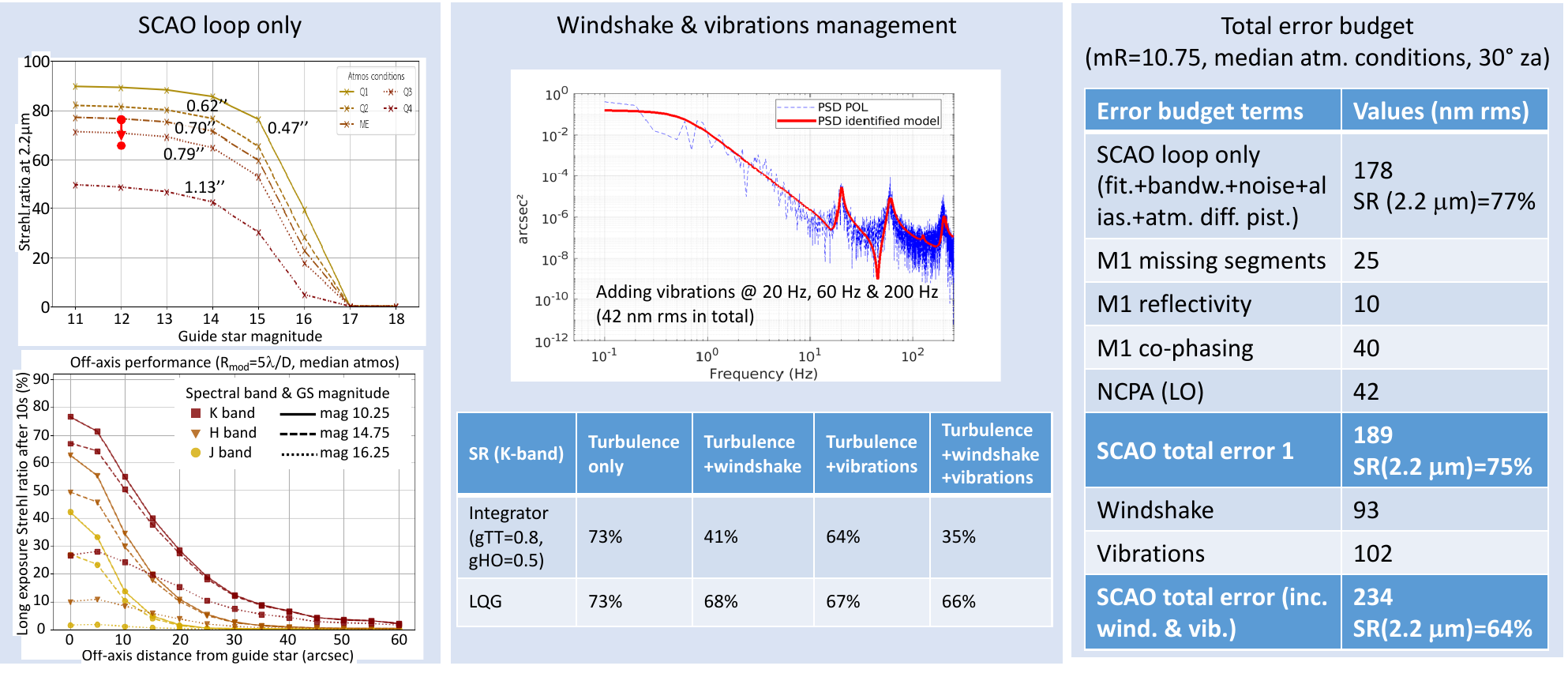} 
\end{center}
\caption{\label{fig:scaoperf} The left panel features at the top the "SCAO loop only" performance (i.e. including fitting, bandwidth, noise propagation, aliasing, atmospheric differential pistons error terms) with respect to the guide star R magnitude, at various seeing values. The red dot under the red arrow shows the expected impact of instrumental effects. At the bottom is the impact of the off-axis distance from the guide star for different guide star magnitude and observing wavelength. The middle panel focuses on the management of windshake and vibrations. The top figure shows a simulated pseudo open loop PSD for a tip sequence that includes vibrations at three different frequencies, 20 HZ, 60 Hz, 200 Hz, for a total amplitude of 42 nm rms, on top of the turbulence and windshake, together with the result of the identification. The lower table gives the pure integrator and LQG performance for different considered scenarios. The right panel gives the total SCAO error budget, making use of the results of the left and middle panels.} 
\end{figure}

The MICADO SCAO system is a pyramid-based AO system, sensing the light in the visible from 600 to 960 nm (fed with 100\% of the light in this bandpass). The pyramid pupils are imaged on the CCD220 detector of the ESO ALICE camera in 96$\times$96 pixels (leading to a subaperture size of 0.40 m). The total number of WFS measurements is then made of about 26000 elements. The pupil will be stabilized in position and clocking. The SCAO system will make use of the ELT M4 and M5 mirrors and the number of controlled modes will be from 2 to 4000. The loop will be running at up to 500 Hz. 

The SCAO system will be able to use a natural guide star with a R magnitude ranging from -1.5 (Betelgeuse) to 16. It will also be able to close the loop on extended objects, up to 1 arcsec in diameter (e.g. Solar system satellites) and will support differential tracking up to 100 arcsec/hour. The guide star will be selectable in a 6\arcsec$\times$20\arcsec\ patrol field.

In terms of AO performance, the MICADO SCAO specification is 60\% of Strehl ratio at 2.2 $\mu$m, at 30$^\circ$ from zenith with a m$_R$=12 reference star, under medium seeing (0.702\arcsec), excluding vibrations and low wind effect.  Figure~\ref{fig:scaoperf} shows the curves of "SCAO loop only" performance (accounting for the fitting error, the bandwidth error, the noise propagation error, the aliasing error and the atmospheric differential piston error, estimated from end-to end simulations made with our \href{https://compass.pages.obspm.fr/website/}{COMPASS platform}), the impact and performance when considering windshake and vibrations and when using a LQG controller, and the total error budget.

Finally, for regular tuning of NCPA and aging diagnostics, the SCAO module will include a dedicated calibration/maintenance unit with sources and a low order adapted ALPAO DMX37 mirror.

\begin{figure}[!t]
\begin{center}
 \includegraphics[height=13.4cm]{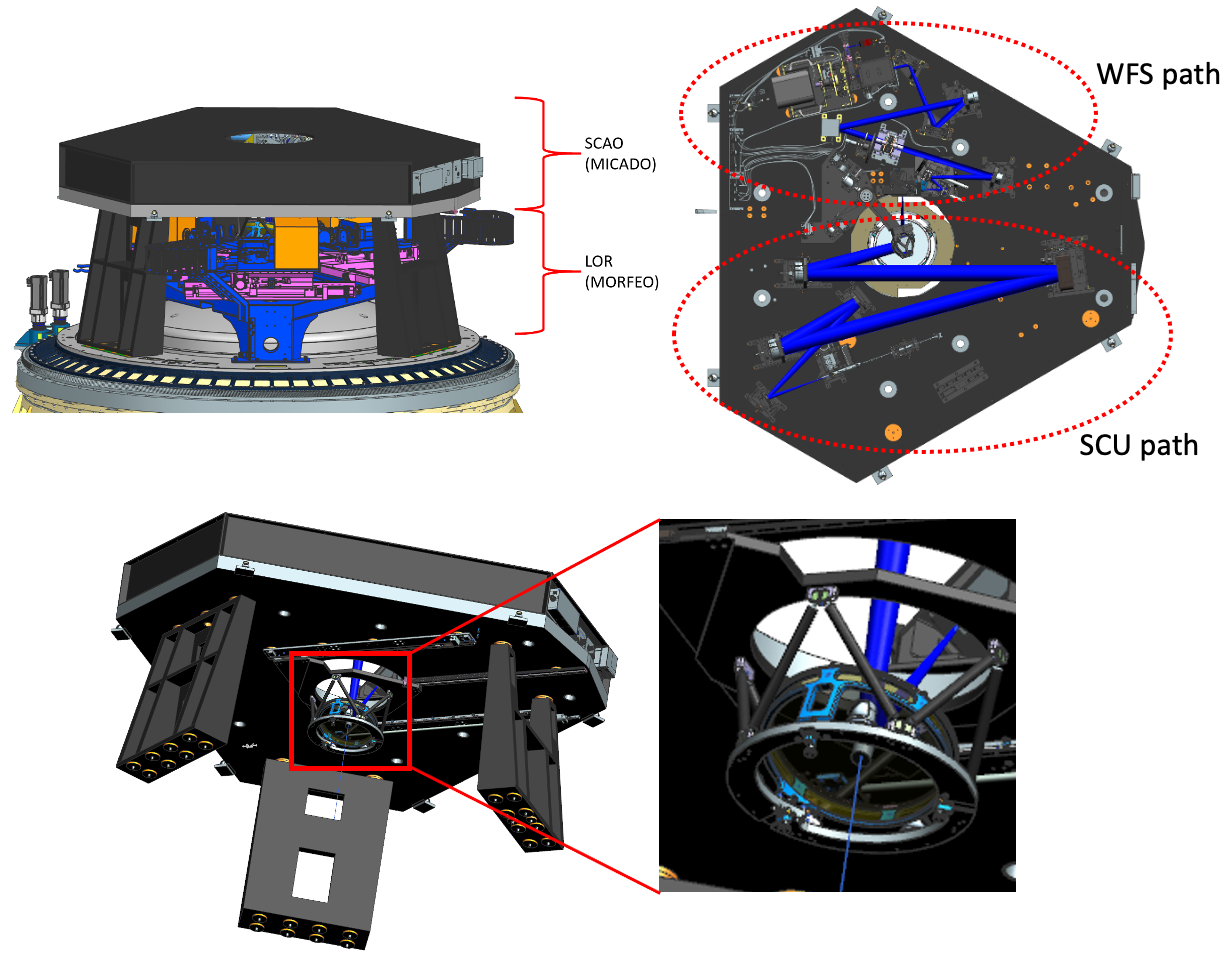} 
\end{center}
\caption{\label{fig:scao-description} Top left: side view of the subsystems inside the so-called Green Doughnut, mounted on top of the MICADO derotator+cryostat, namely the SCAO module at the top and the MORFEO LOR WFS at the bottom. Top right: top view of the SCAO bench in its on-sky configuration, showing the WFS path and the SCU path. Bottom: bottom views of the SCAO bench, showing the dichroic plate assembly.}
\end{figure}

The main sub-systems of the SCAO module are: a pyramid-based wavefront sensor, a calibration unit (so-called SCU), a dichroic plate assembly, a support structure, a real-time computer, a control software. Figure~\ref{fig:scao-description} shows different views of the SCAO module and these different sub-systems. The analysis made to build the optical design of the SCAO WFS is presented in these proceedings by Cohen et al. (2024)\cite{cohen24}. 

\newpage

\begin{figure}[!t]
\begin{center}
   \begin{tabular}{c c c c}
 \includegraphics[width=3.2cm]{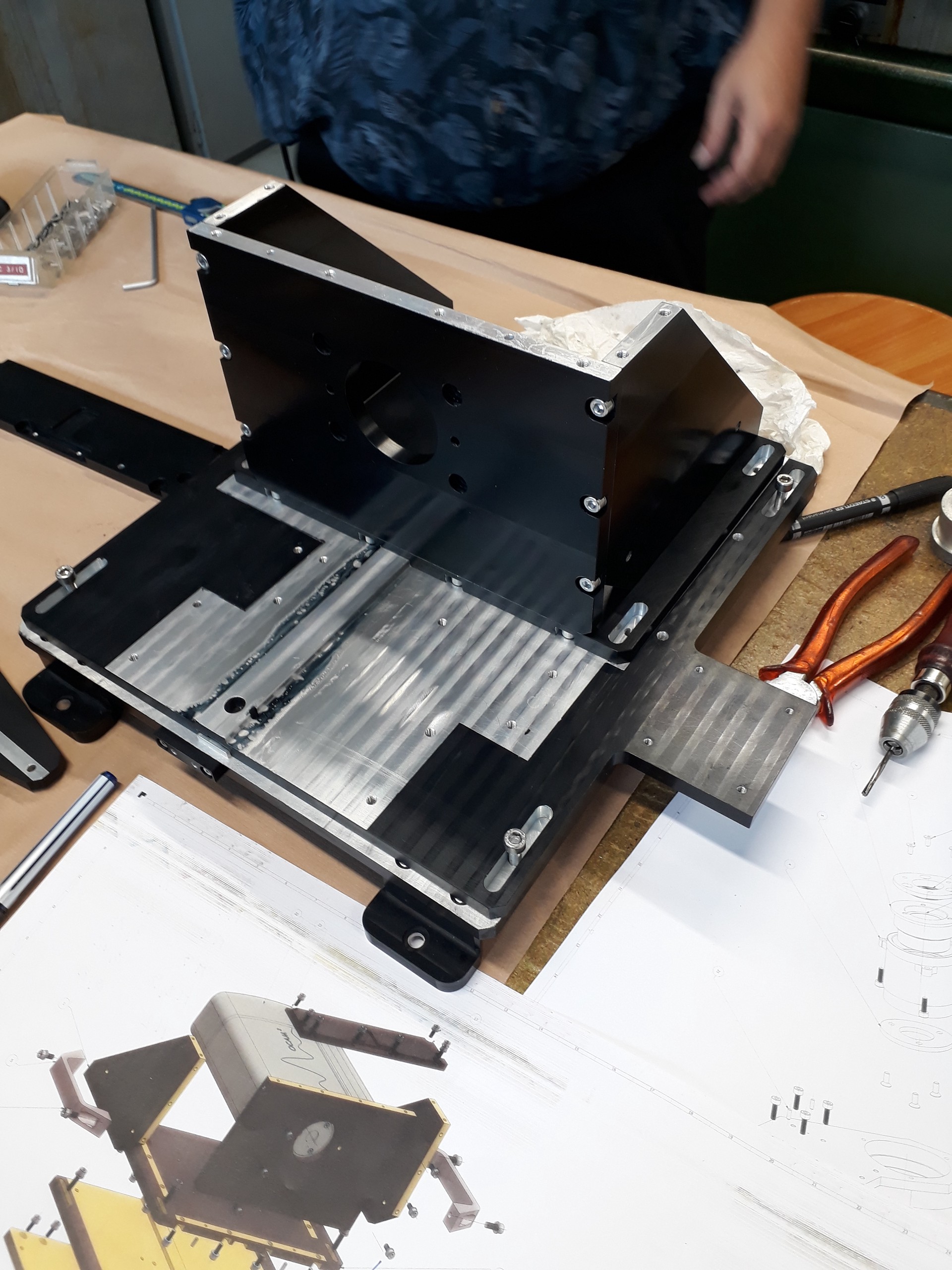}  & \includegraphics[width=3.2cm]{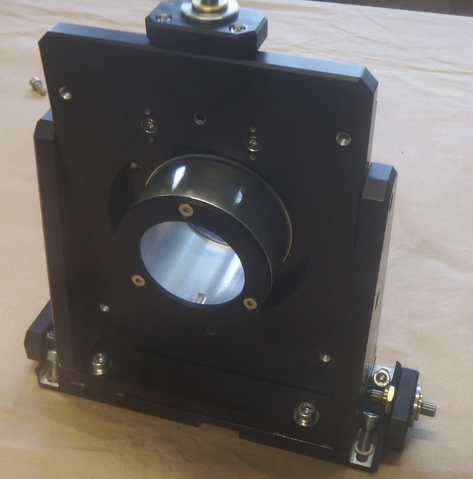} & \includegraphics[width=3.2cm]{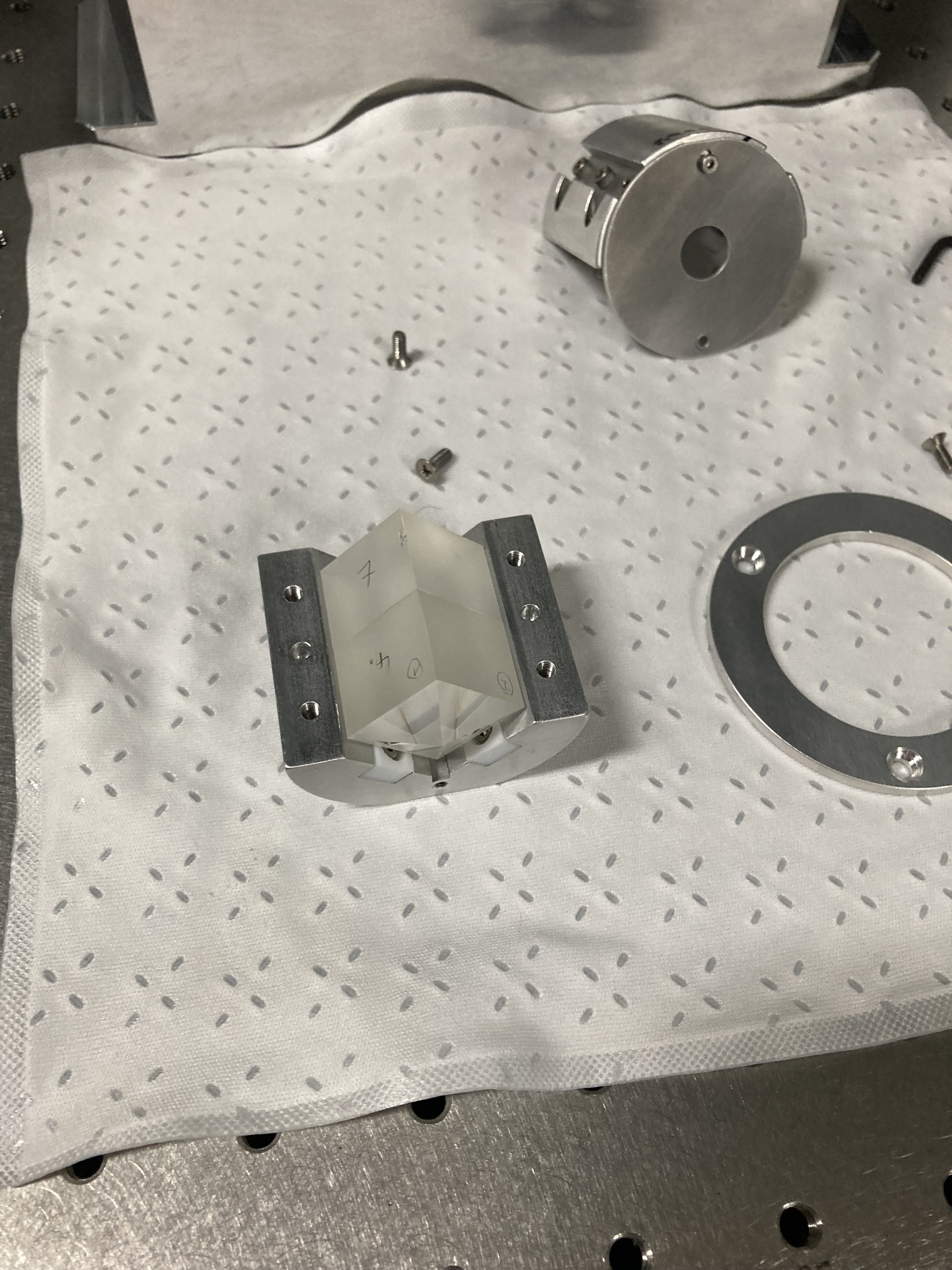} & \includegraphics[width=3.2cm]{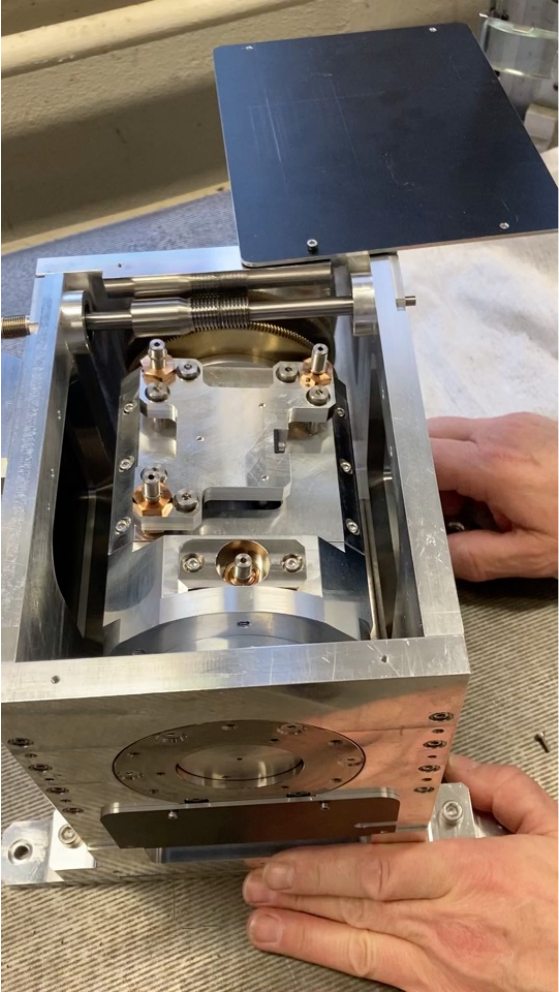} \\
  \includegraphics[width=3.2cm]{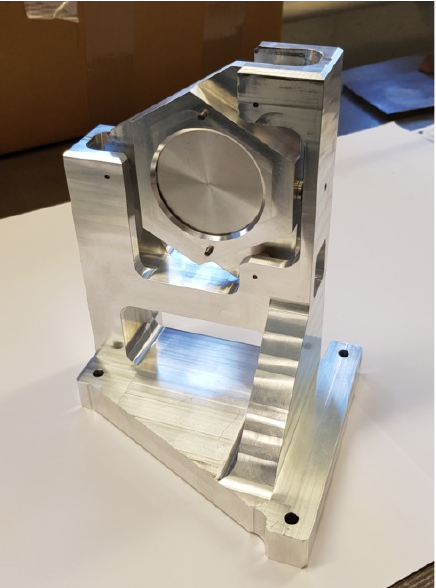} & \includegraphics[width=3.2cm]{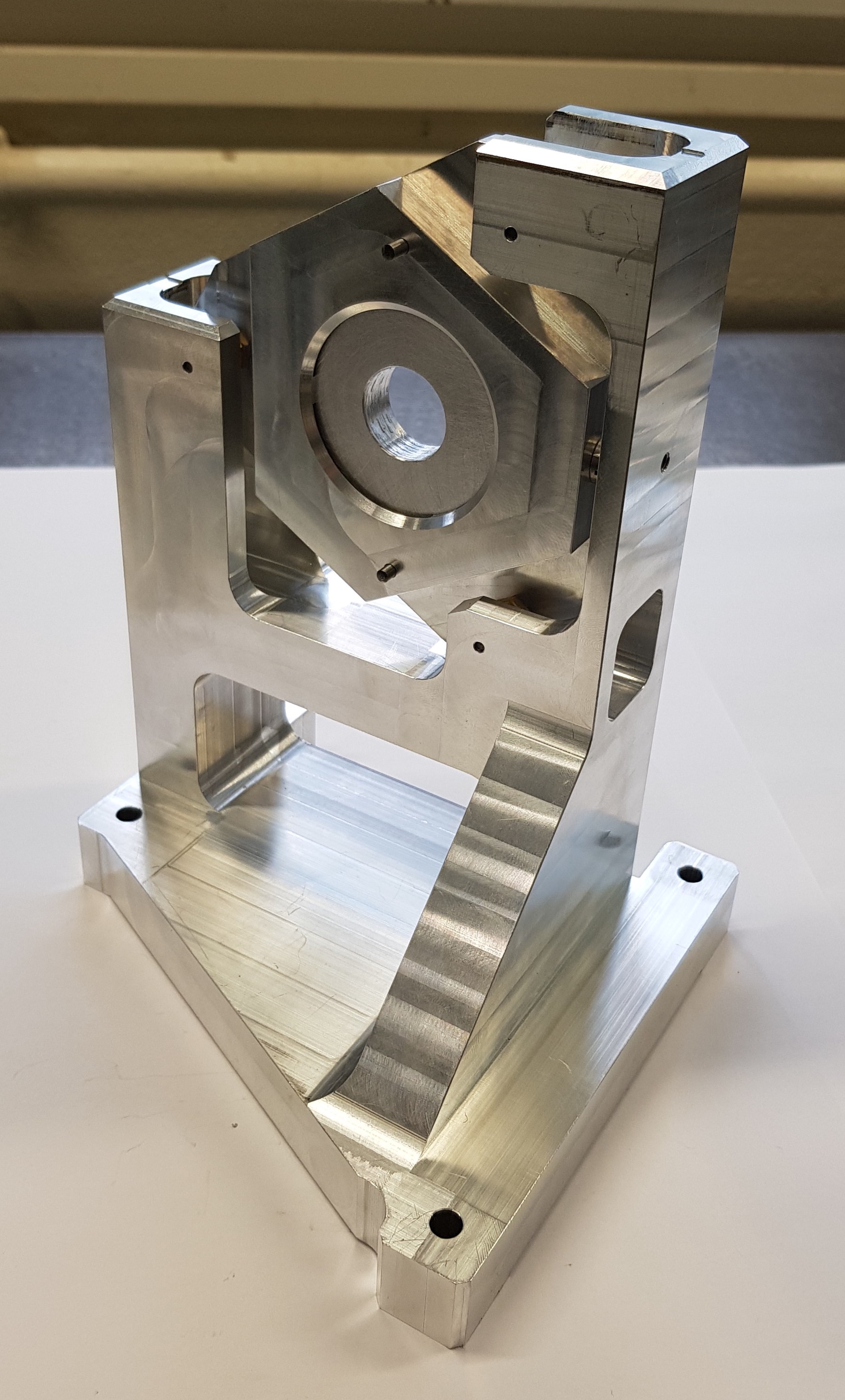} & \includegraphics[width=3.2cm]{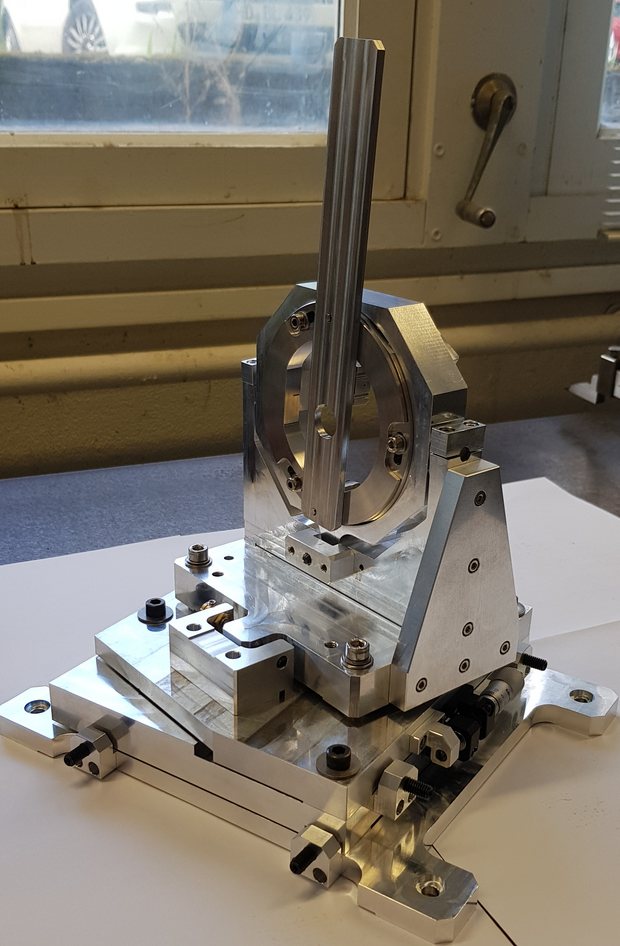} & \includegraphics[width=3.2cm]{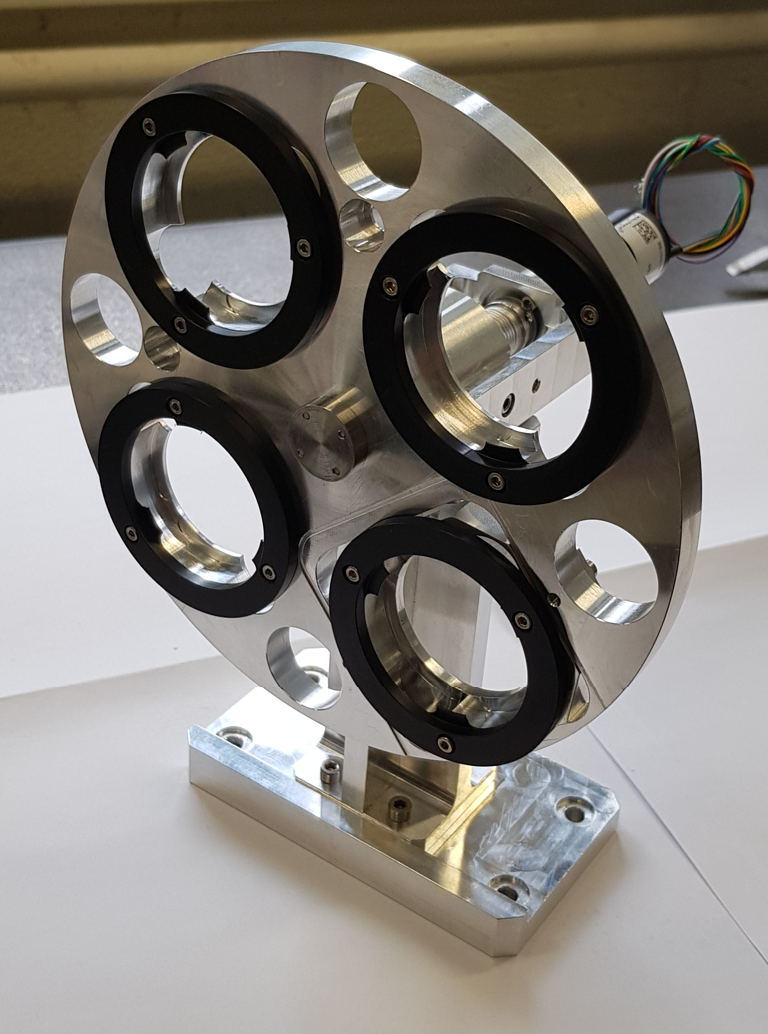} \\
  \includegraphics[width=3.2cm]{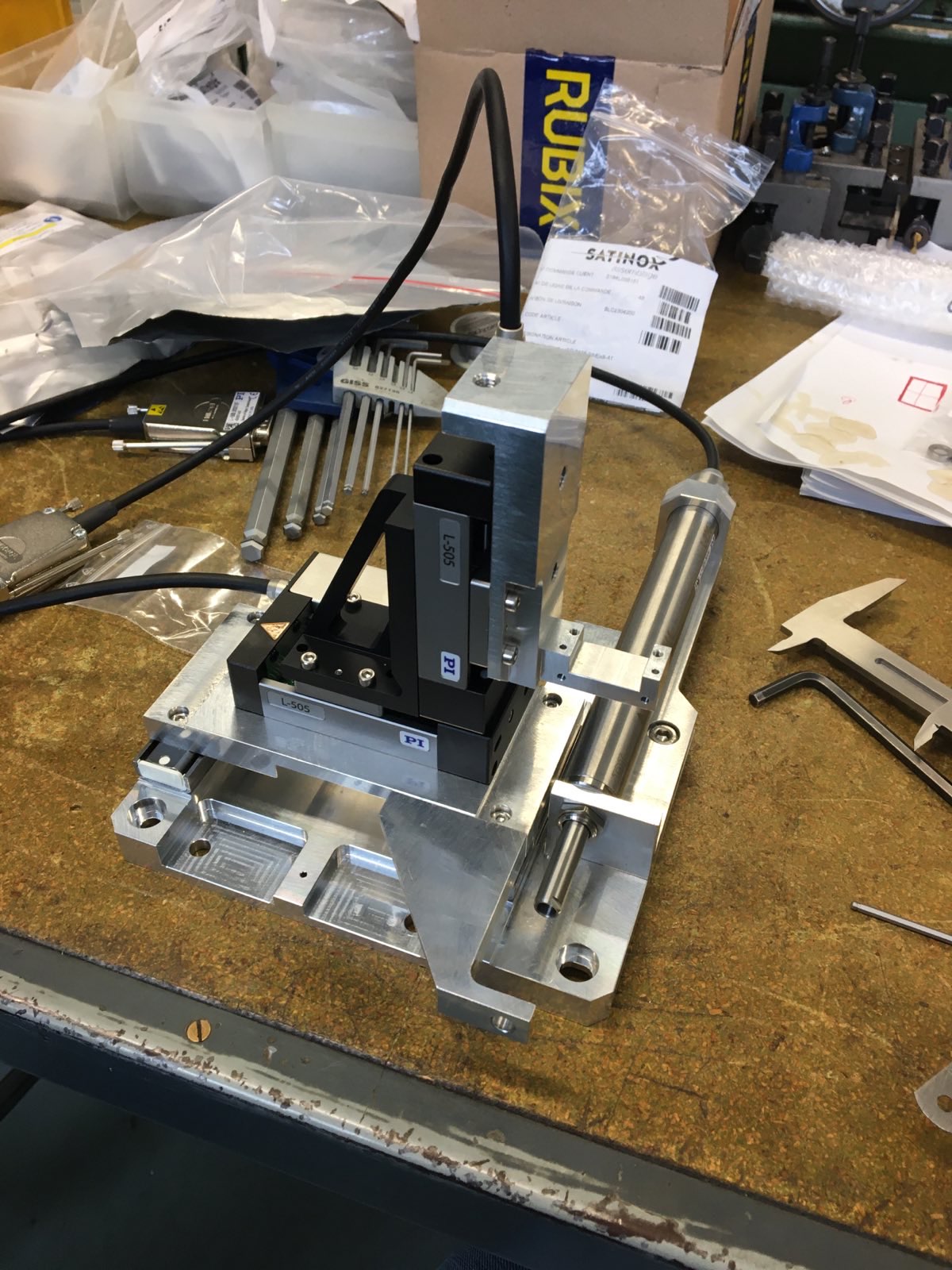} & \includegraphics[width=3.2cm]{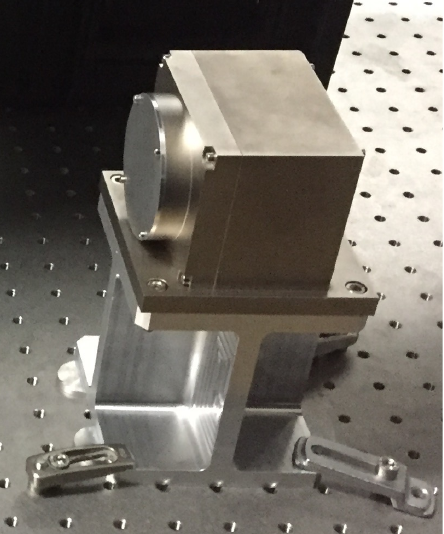} & \includegraphics[width=3.2cm]{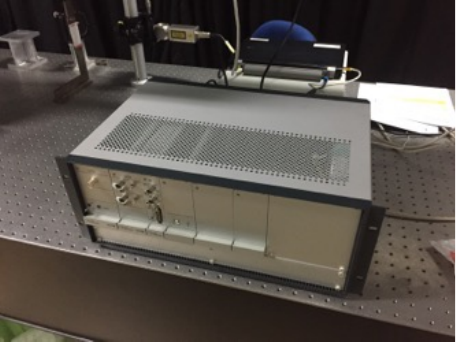} & \includegraphics[width=3.2cm]{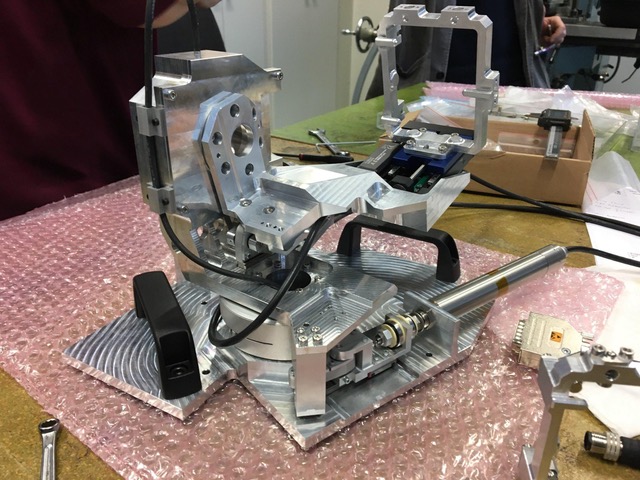} \\
 \includegraphics[width=3.2cm]{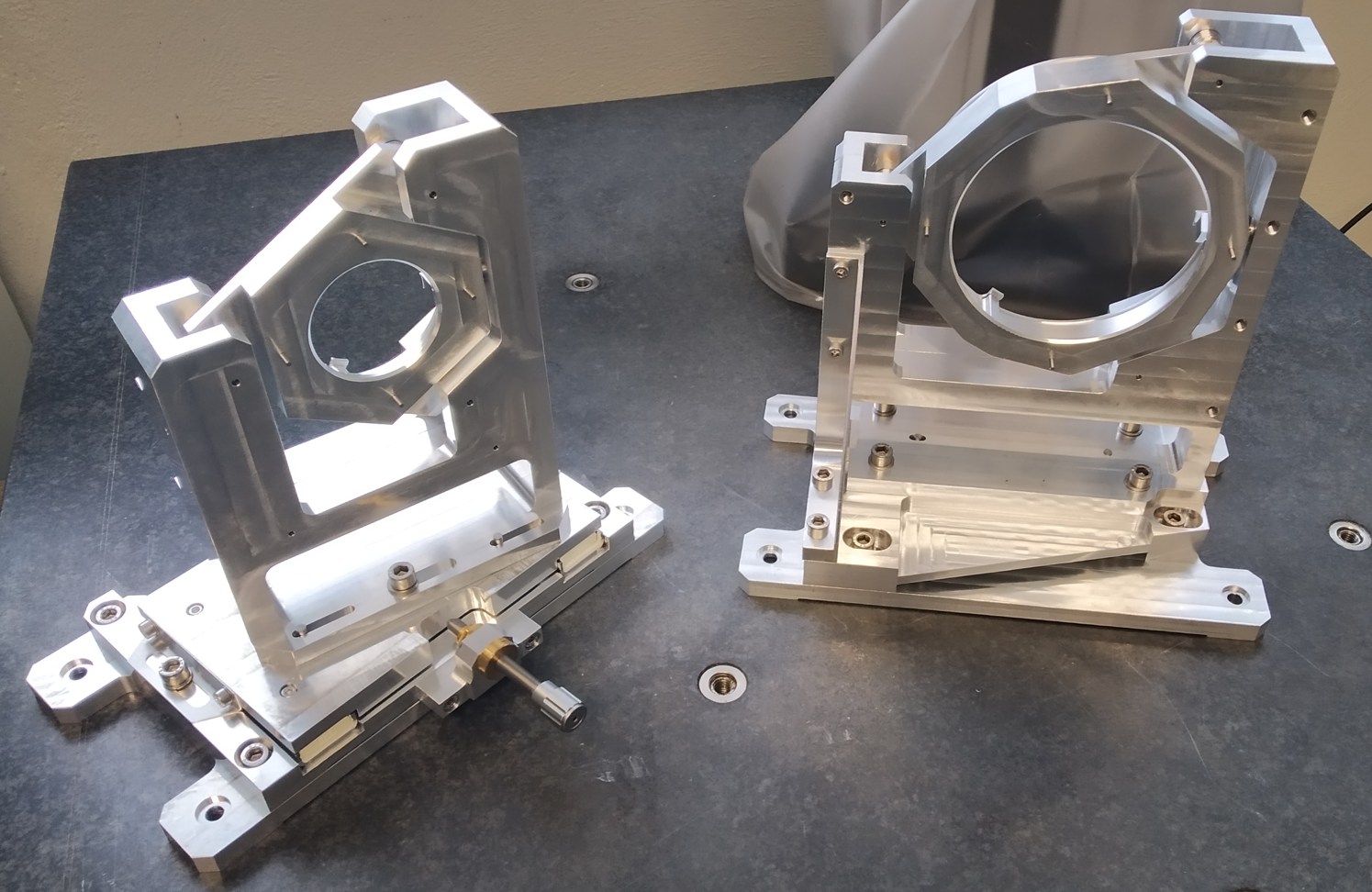} & \includegraphics[width=3.2cm]{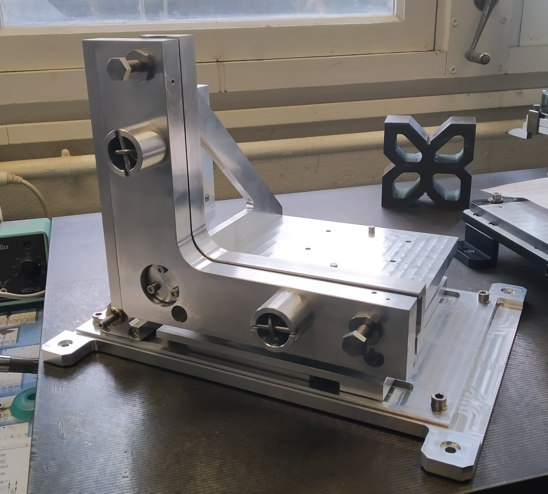} & \includegraphics[width=3.2cm]{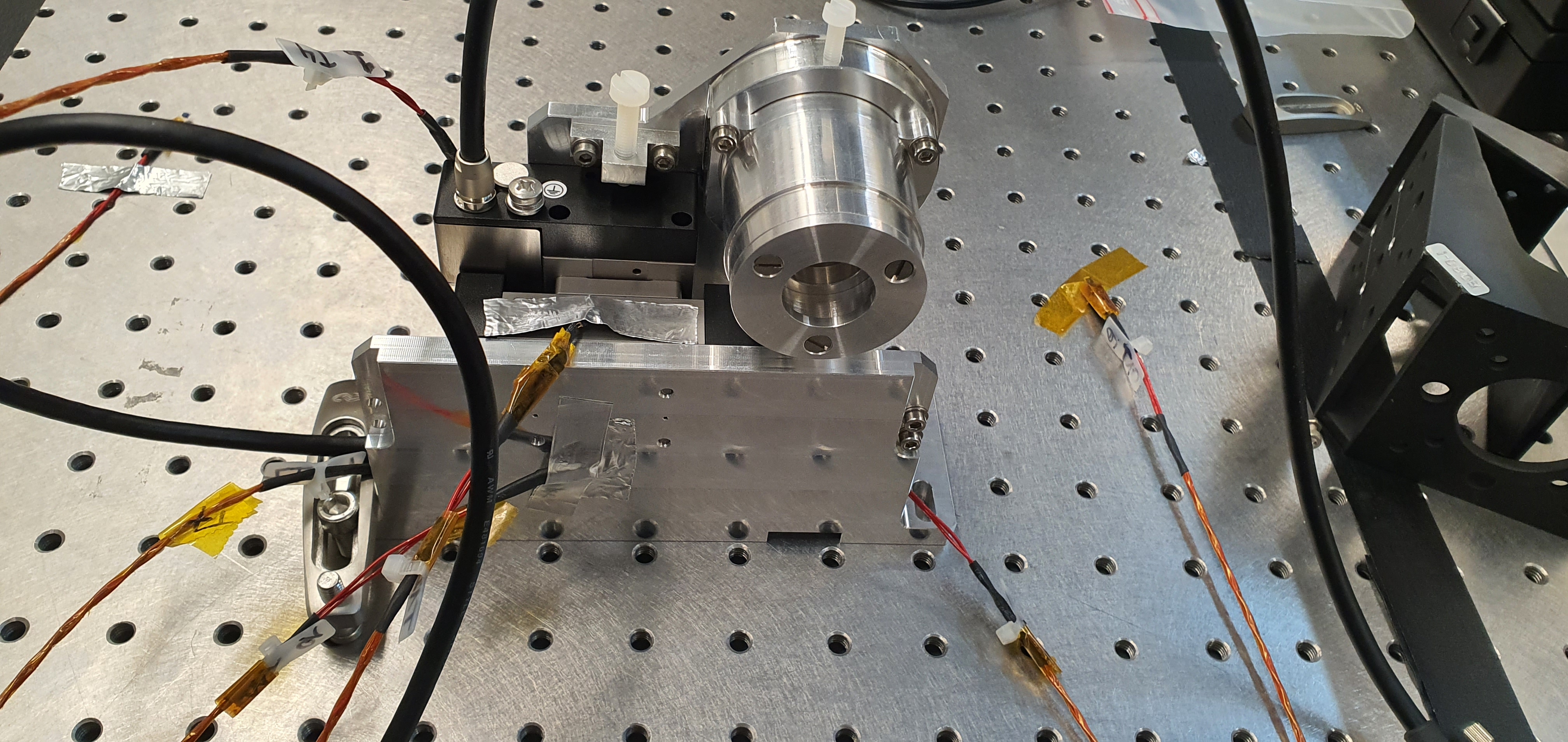} & \\
  \end{tabular}
\end{center}
\caption{\label{fig:meca} Delivered/manufactured subsystems, from top to bottom and left to right: WFS camera support, pyramid prism support assembly, pyramid prism support, K-mirror, WFS folding mirrors supports, WFS off axis parabolas support, NCPA phase screens support, SCU source selection module, modulation system and electronics, field selector, SCU folding mirrors supports, common DMs support, X/Y pupil management subsystem. }
\end{figure}

\section{Delivered/manufactured subsystems}

This section focuses on and illustrates the various mechanical assemblies that have been already manufactured or delivered. They are the following (Fig.~\ref{fig:meca}):
\begin{itemize}
\item the WFS camera support, whose mechanics has been manufactured, blackened and assembled. This support is compatible with both the final WFS camera ALICE, developed by ESO and whose delivery is not expected before Feb. 2025, and the SCAO AIT camera, an FLI OCAM2k, chosen to mitigate the late delivery of ALICE in the AIT planning.
\item the X/Y pupil management subsystem, whose mechanics has been manufactured, blackened and assembled. The PI L505 linear stages have been delivered and their control through Beckhoff terminals already fined tuned. A dedicated "special device" software component following the ESO Instrument Control Software Framework has been developed to control this subsystem and the full testing of the subsystem is foreseen during summer 2024.
\item the pyramid prism support, that has been manufactured, blackened and assembled. 
\item the K-mirror assembly, that has been manufactured, assembled and is being blackened. Developed to control the pupil in clocking, its critical specifications in terms alignment and stability have been already demonstrated\cite{clenet03}.
\item the WFS folding mirrors supports, that have been manufactured, assembled and are being blackened
\item the WFS off-axis parabolas supports, that have been manufactured, assembled and are about to be blackened
\item the modulation system, manufactured together with its electronics by CEDRAT, that has very recently been tested and validated
\item the NCPA phase screens support wheel, that has been manufactured, assembled and is about to be blackened. It aims at handling the fixed part of the NCPA (coming from MICADO and the WFS optics)
\item the field selector, that has been manufactured, assembled and is about to be blackened. The PI L220 stages used for its mirrors' displacement have been delivered to us and their control through Beckhoff terminal recently fine-tuned. A full testing of this subsystem is foreseen this summer
\item the SCU source selection module, that has been manufactured , assembled and is about to be blackened. The PI L220 and L505 used for the sources' alignement have been delivered to us and their control through Beckhoff terminal recently fine-tuned 
\item the SCU folding mirrors supports, that have been manufactured, assembled and are being blackened
\item the common support for the low order calibration and high order AIT DMs, that has been manufactured and assembled
\end{itemize}

Hence, the mechanical sub-systems that remain to be manufactured are the WFS pupil viewer assembly, the WFS ADC, the WFS field lens support, the WFS pupil steering mirror, the SCU deployment arm, the SCU off axis parabolas supports, the SCAO carbon bench/feet and the SCAO dichroic support.

\section{Status of optics delivery}
\begin{figure}[!t]
\begin{center}
   \begin{tabular}{c c}
 \includegraphics[width=7cm]{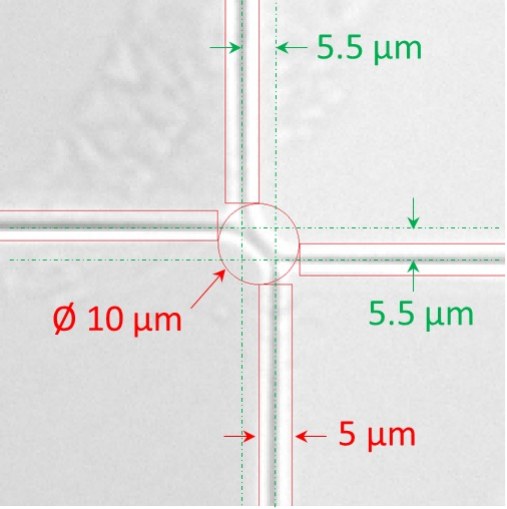}  & \includegraphics[width=7cm]{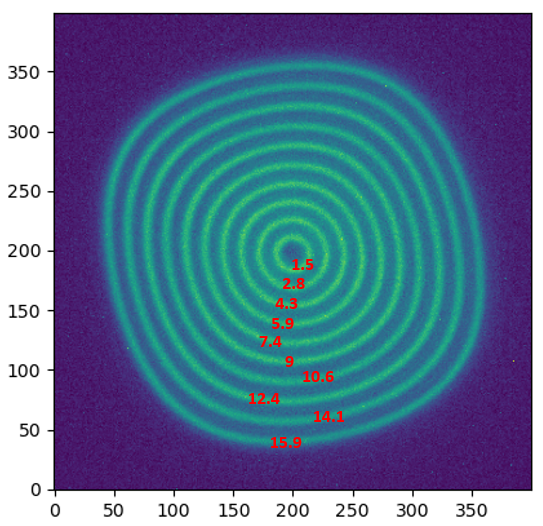}\\
  \includegraphics[width=7cm]{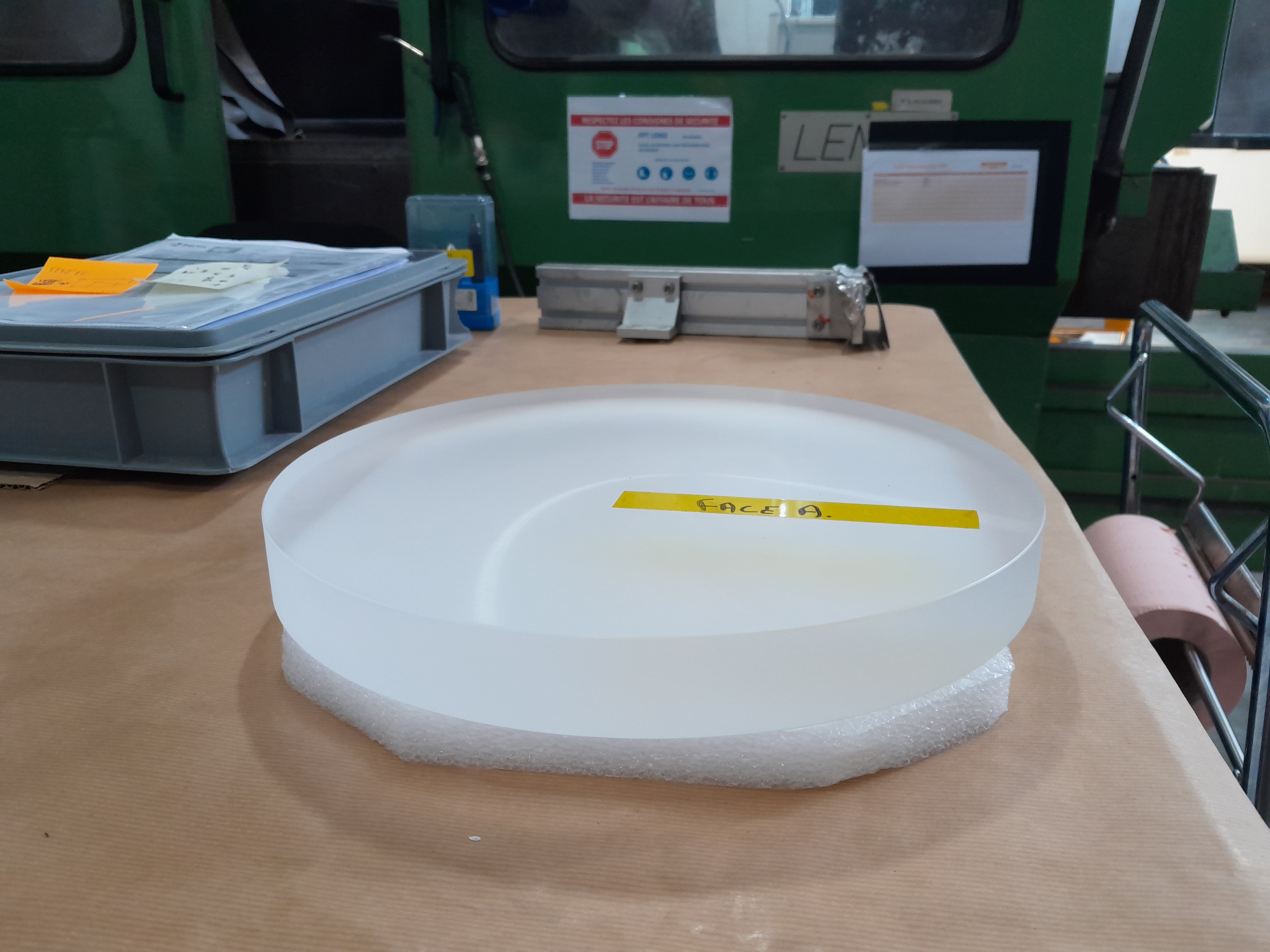}  & \includegraphics[width=8cm]{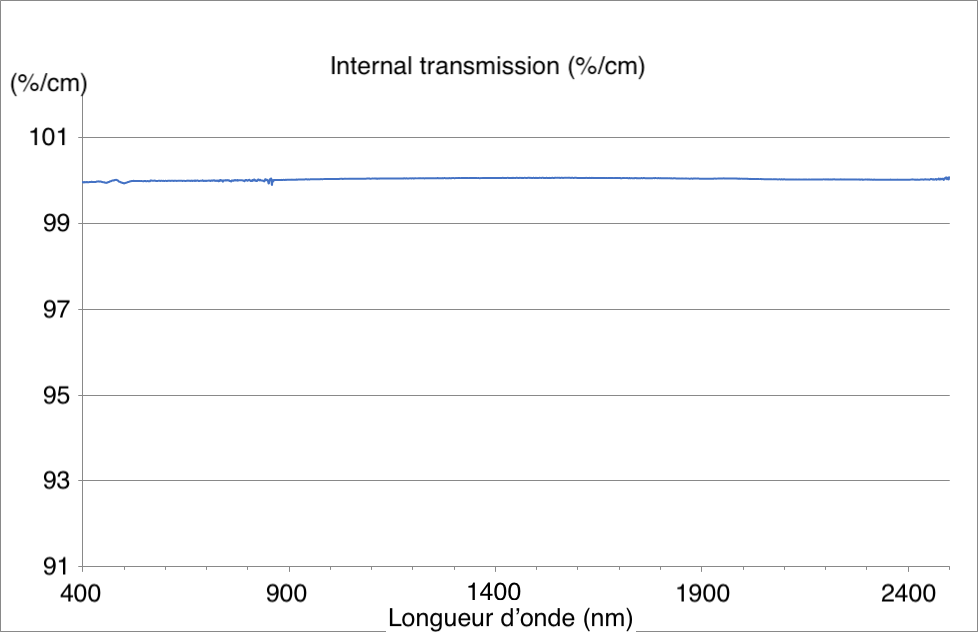}\\
 \end{tabular}
\end{center}
\caption{\label{fig:demo-optics} Top left: Measurements of pyramid edges width and vertex size on one of the front pyramid prism. Top right: modulation traces for different modulation radius in $\lambda/D$. Bottom left: Photo of the first blank delivered by Hellma to Bertin, in Bertin's premises. Bottom right: measurement by Hellma of the internal transmission of the first blank}
\end{figure}

Several optical parts have already been delivered: the modulation mirror, the neutral NCPA phase screens, the field selector mirrors, the pyramid double-prisms, the WFS pupil viewer folding mirror, the AIT ALPAO high order (64$\times$64) deformable mirror and all the optics (folding mirrors, off axis parabolas, ELT pupil masks, turbulence phase screens) for the SCU extension module that simulates the adaptive telescope and the turbulence. Figure~\ref{fig:demo-optics} shows validation illustrations of the modulation system (modulation traces for different modulation radius in $\lambda/D$) and the pyramid double-prism (measurement of the pyramid edges width and vertex size).

Most of the remaining optical parts have been ordered and are pending delivery:
\begin{itemize}
\item reflective optics, subject to a contract placed to Bertin and whose delivery is expected in Oct. 2024: WFS pupil steering mirror, WFS off axis parabolas, WFS folding mirrors, K-mirror mirrors, SCU off-axis parabolas, SCU folding mirrors, SCU deployment arm mirror
\item refractive optics, subject to a contract placed to Bertin and whose delivery is expected in Q1 2025: WFS field lens, WFS ADC prisms, WFS beam splitter, WFS pupil imaging lenses, WFS pupil viewer path pupil imaging lenses
\item dichroic optics (Fig.\ref{fig:demo-optics}), made of a 305 mm$\times$30mm CaF2 plate. The raw material of the first blank has been delivered by Hellma to Bertin. Delivery from Bertin to our colleagues from LMA for the dichroic coating of this first blank is planned for late Nov. 2024. A second blank will be delivered by Hellma to Bertin in Oct. 2024. After polishing it will be delivered to LMA and kept without further treatment as a spare.
\item SCU calibration low order DM, ordered to ALPAO, will be delivered in Q4 2024
\end{itemize}

\newpage

\section{RTC development}
Following the RTC architecture required by ESO for ELT instruments, the MICADO SCAO RTC is divided into a hard real time core (HRTC), a soft real time cluster (SRTC) and a RTC communication infrastructure.

\begin{figure}[!t]
\begin{center}
   \begin{tabular}{c}
  \includegraphics[width=13cm]{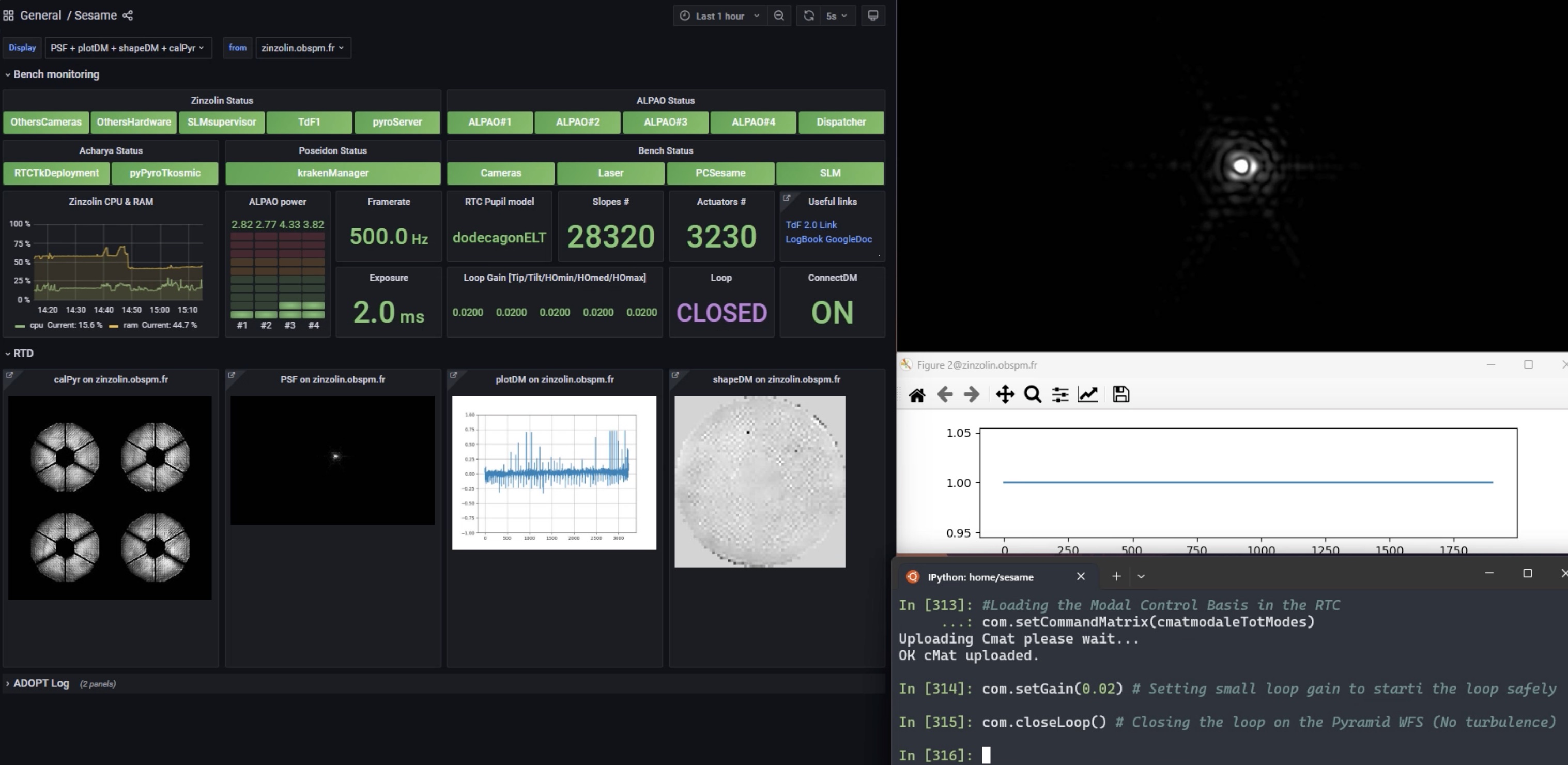}  \\ \includegraphics[width=13cm]{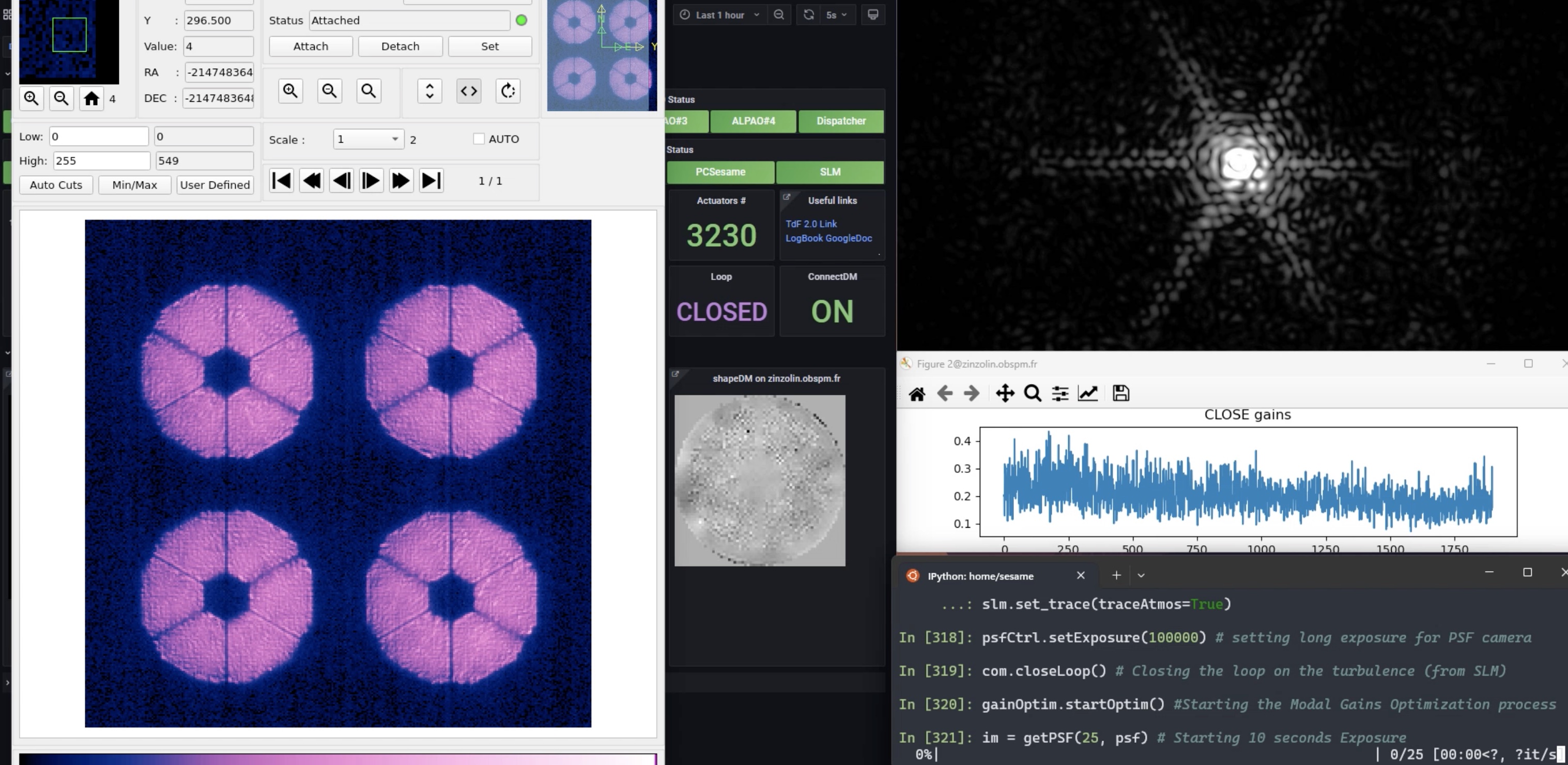}\\
  \end{tabular}
\end{center}
\caption{\label{fig:close-loop} Control panel of the SCAO loop on our AO R\&D S\'esame bench for the first close loop on bench at ELT scale. Top: closing the loop with no turbulence. Bottom: closing the loop with turbulence.}
\end{figure}

The HRTC is based on the COSMIC platform\cite{ferreira24}. Today, the hard real-time pipeline has been developed and heavily tested(Fig.~\ref{fig:close-loop}), except the LQG part, which is subject to a dedicated R\&D between LESIA and LCF. 

The SRTC is relying on the ESO RTC toolkit. We pro-actively integrate every update of this SW package, using several nodes.

A full loopback between HRTC and SRTC has been already tested, including the background computation and the command application tasks.

We have deployed two separated RTC set-up\cite{sevin24}. One full ELT scale set up, with the different nodes required by the ESO ELT RTC architecture. It allows the integrate and test this required architecture. A second one is deployed on our AO R\&D S\'esame bench, to test the HRTC algorithms and performance. Hence, we have been able to run a stable loop at 500 Hz using this MICADO SCAO RTC (COSMIC HRTC pipeline and RTC toolkit SRTC) with the MICADO double pyramid prism and our 64$\times$64 high order AIT DM. To our knowledge it is the first close loop on bench at ELT scale performed in the community ! 

\section{First AIT configuration: the $\beta$ flat configuration}
\begin{figure}[t]
\begin{center}
 \includegraphics[height=10cm]{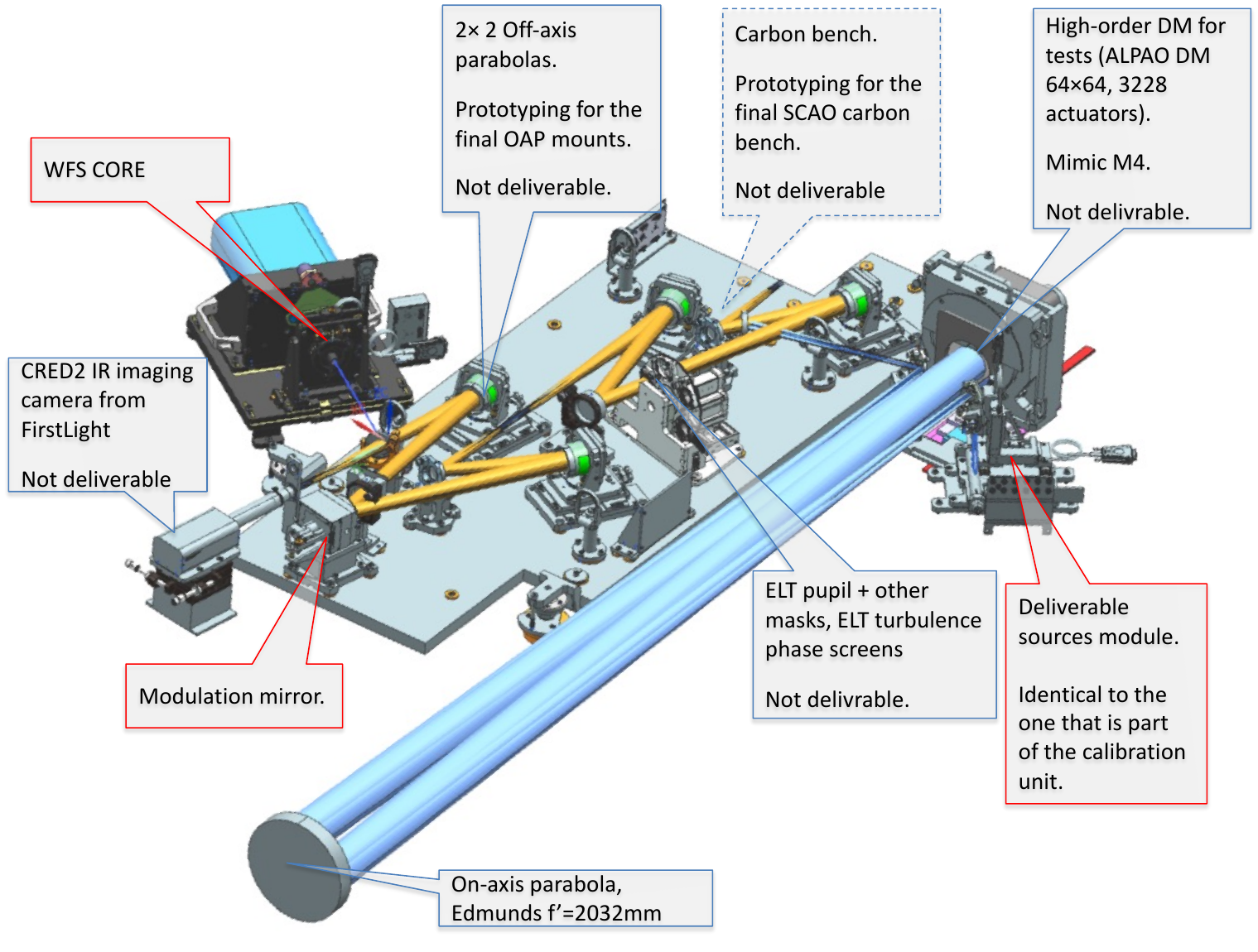} 
\end{center}
\caption{\label{fig:betaflat} Description of the $\beta$ flat configuration.}
\end{figure}

This  AIT set up is made of:
\begin{itemize}
\item the least number of final SCAO sub-systems required to close the loop:
\begin{itemize}
\item the pyramid assembly
\item the modulation assembly
\item the source module
\end{itemize}
\item the SCU AIT module, that allows to simulate the telescope and the atmosphere, comprising:
\begin{itemize}
\item the AIT carbon bench (allowing prototyping the final SCAO carbon bench)
\item the AIT HO DM (64x64 ALPAO DM)
\item the pupil module assembly (ELT pupil mask+turbulence phase screens)
\end{itemize}
\item the AIT electronics cabinets with the required (final) Beckhoff terminals
\end{itemize}

We are currently finishing the integration of this $\beta$ flat configuration. What remains to be done is:
\begin{itemize}
\item to integrate the Beckhoff terminals  in a just delivered AIT electronics cabinet
\item to move parts to the AIT room from our R\&D S\'esame room
\item to integrate them together with the SCU AIT module
\end{itemize}

We expect to start the operations in this configuration this September, with the aim to perform the tests and validation of the various AO/RTC algorithms in a polychromatic configuration and to perform the tests and validation of few subsystems control software (low level + middleware). It will take place in a dedicated AIT room at Observatoire de Meudon, named "ATLAS"

\section{Next AIT phases in France}
\begin{figure}[t]
\begin{center}
 \includegraphics[height=8cm]{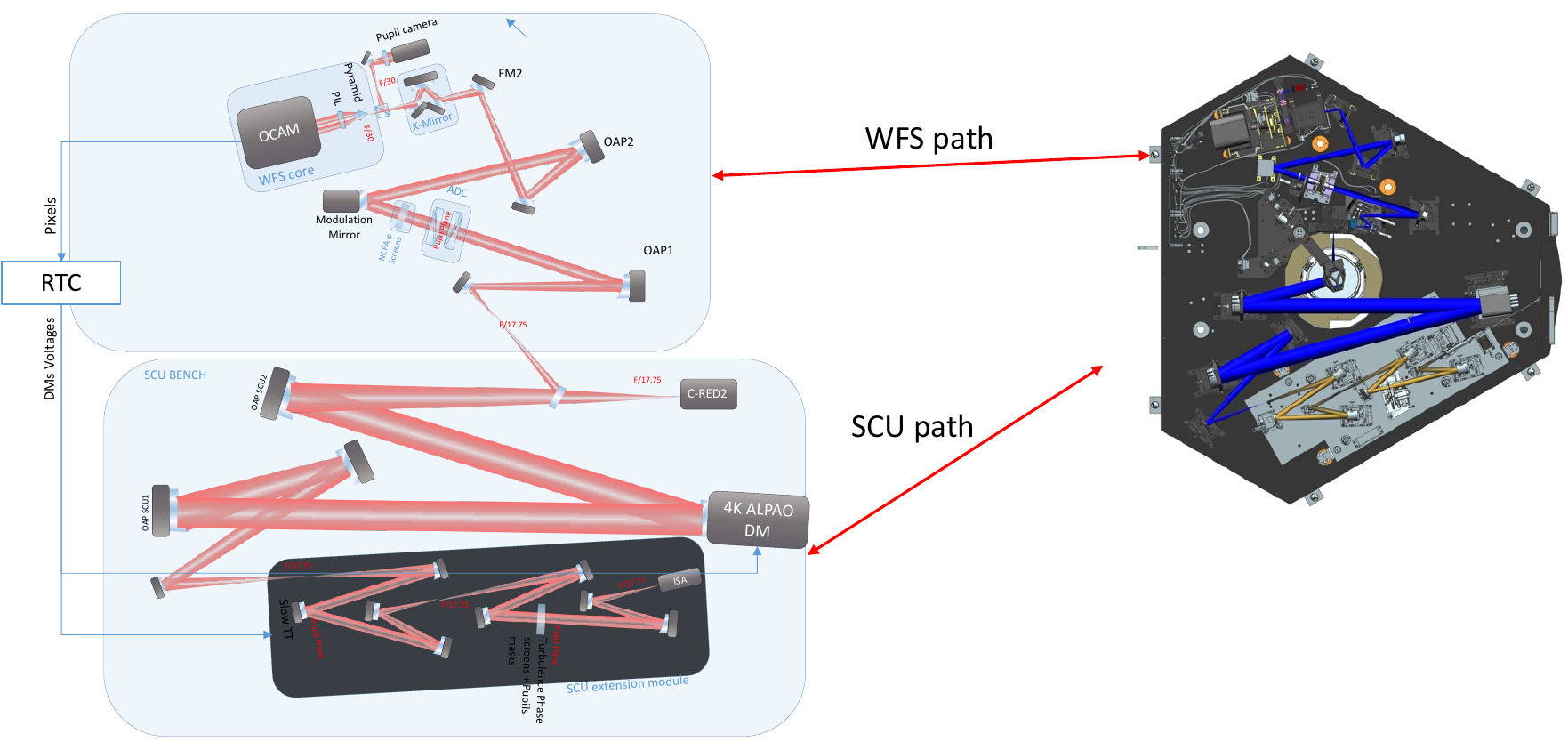} 
\end{center}
\caption{\label{fig:ait-conf} The AIT flat configuration (left) and  the final AIT configuration (right).}
\end{figure}

\begin{figure}[!h]
\begin{center}
   \begin{tabular}{c c}
 \includegraphics[width=7.5cm]{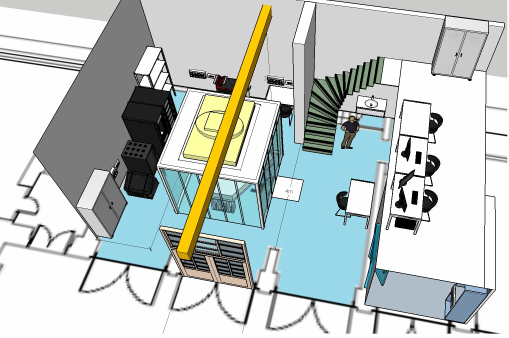}  & \includegraphics[width=8.5cm]{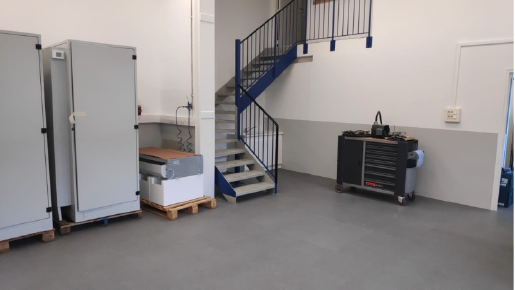}\\
 \end{tabular}
\end{center}
\caption{\label{fig:communs} Left: Drawing of the final AIT room in Observatoire de Meudon. Right: Photo of this AIT room after the end of the refurbishing and in use for the AIT of the Gravity+ RTC cabinets}
\end{figure}

After the $\beta$ flat configuration, the next phase of the SCAO MAIT plan is the so-called flat configuration where the WFS and the SCU are integrated on separated benches with all their elements (Fig.~\ref{fig:ait-conf}, left). It will still be in the ATLAS AIT room. It will allow to test the functionalities of all the components non present in the $\beta$ flat configuration. The SCAO elements will then be integrated on the final SCAO bench, in its final AIT configuration, i.e. with the elements already available in the $\beta$ flat configuration that allow to simulate the telescope and the atmosphere (Fig.~\ref{fig:ait-conf}, right). It will occur in a dedicated AIT room in Observatoire de Meudon, that has been recently completely refurbished for the MICADO SCAO AIT  (Fig.~\ref{fig:communs}). This AIT configuration will allow the validation of the SCAO module specifications and will be used for a partial MICADO SCAO PAE. After this partial PAE, the specific MAIT elements will be removed to get the final on-sky version of the SCAO module.

\acknowledgments 
This work has benefited from the support of 1) the French Programme d'Investissement d'Avenir through the project F-CELT ANR-21-ESRE-0008, 2) 2) the CNRS 80 PRIME program, 3) the CNRS INSU IR budget, 4) the Action Sp\'ecifique Haute R\'esolution Angulaire (ASHRA) of CNRS/INSU co-funded by CNES, 5), the Observatoire de Paris and 6) the Ile de France region (DIM ACAV/ACAV+ and ORIGINES).

\bibliography{report} 
\bibliographystyle{spiebib} 

\end{document}